# A Multi-Smartwatch System for Assessing Speech Characteristics of People with Dysarthria in Group Settings

Harishchandra Dubey[#], J. Cody Goldberg, Kunal Mankodiya[*]
Dept. of Electrical, Computer, and Biomedical Engineering
University of Rhode Island,
Kingston, RI 02881, USA
dubey@ele.uri.edu[#], kunalm@uri.edu[*]

Leslie Mahler[+]
Dept. of Communicative Disorders
University of Rhode Island,
Kingston, RI 02881, USA
lmahler@uri.edu[+]

*Abstract*—Speech-language pathologists (SLPs) frequently use vocal exercises in the treatment of patients with speech disorders. Patients receive treatment in a clinical setting and need to practice outside of the clinical setting to generalize speech goals to functional communication. In this paper, we describe the development of technology that captures mixed speech signals in a group setting and allows the SLP to analyze the speech signals relative to treatment goals. The mixed speech signals are blindly separated into individual signals that are preprocessed before computation of loudness, pitch, shimmer, jitter, semitone standard deviation and sharpness. The proposed method has been previously validated on data obtained from clinical trials of people with Parkinson disease and healthy controls.

## I. Introduction

Approximately 7.5 million people in the United States have speech disorders that can be caused by a number of neurological diagnoses. *Dysarthria* refers to speech disorders resulting from abnormalities in the speed, strength, range of motion, steadiness, or accuracy of movement of muscles needed for speech production. Speech-language pathologists (SLPs) are involved in the evaluation, diagnosis, and treatment of people with dysarthria. Dysarthria can be caused by progressive neurological disorders such as Parkinson's disease (PD) [1]-[5], congenital neurological disorders such as cerebral palsy (CP) [6]-[9] or Down syndrome (DS) [10]-[12], and acquired neurological disorders such as stroke [13]-[15] and traumatic brain injury (TBI) [16]-[17]. Speech-language pathologists (SLPs) frequently use vocal exercises in the treatment of patients with speech disorders such as dysarthria. Patients receive treatment in a clinical setting and need to practice outside of the clinical setting to generalize speech goals to functional communication. Deploying remote treatment strategies in home settings and monitoring speech characteristics in social scenarios is challenging due to inherent limitations. *EchoWear* is a wearable speech monitoring system designed to provide a tool for patients as well as SLPs to assess speech outside of the clinical environment, in an ecologically valid communication situation. It allows the remote collection of speech data with a smartwatch worn by the patients. *EchoWear* analyzes the speech data to compute loudness, pitch, measures of perturbation, and frequency of practice, that can be accessed by SLPs for assessing speech characteristics of individual patients. It reduces the logistic requirements and cognitive demands on patients while increasing accuracy and frequency of speech exercises. Recently, increasing numbers of SLPs have adopted telerehabilitation services to improve delivery of speech treatments. Our pilot studies provided evidence of the efficacy and accuracy of *EchoWear* for reliable collection of speech data and its use for calculating loudness and pitch for individuals with PD as well as healthy controls [18]. It is important to monitor the accuracy and frequency of speech exercises outside of the clinic environment but this can be challenging. One challenge is collecting data from an individual patient when there are multiple speakers. The purpose of this study was to assess *EchoWear* in a group environment to evaluate the reliability and validity of speech data collection in the context of multiple speakers. We designed a multi-smartwatch system that combines the blind source separation with noise reduction to provide access to reliable estimates of individual patient's speech signal to address this challenge. The preprocessed speech signal was used to compute speech features of loudness, pitch, shimmer, jitter, sharpness and semitone standard deviation that can be used for assessment of perceptual speech quality. This paper makes the following contributions:

a) Application of the multi-smartwatch system for capturing pathological speech during speech exercises in a group setting;
b) Investigating speech characteristics of people with dysarthria in a group setting;
c) Using independent component analysis (ICA) for the separation of mixed signals acquired simultaneously from speech exercises in a group ;
d) Experimental validation of computing speech quality metrics from disordered speech signals;
e) The individual speech signals extracted from the mixtures can be used by SLPs or patients to monitor or track speech quality over a period of time. The individual speech signals extracted from the mixtures can be used by SLPs or

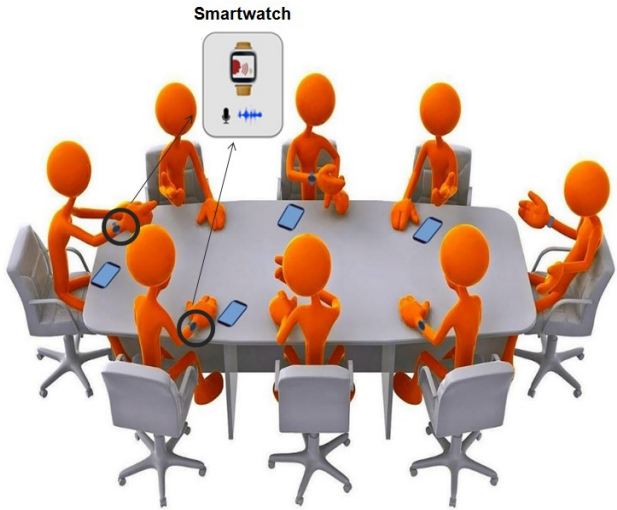

Fig. 1. Acoustic scene for capturing the mixed signals from group settings.

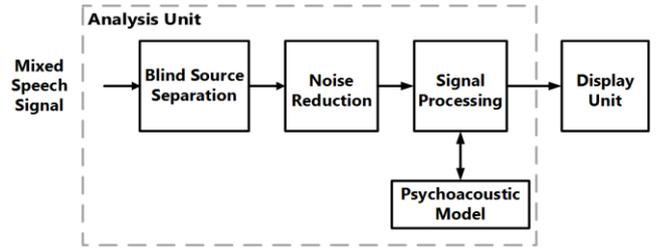

Fig. 2. Speech processing chain for group scenarios.

*patients to monitor or track speech quality over a period of time.*

## II. MULTI-SMARTWATCH SYSTEM

### A. Wearable Acoustic Sensing and Retrieval

The smartwatch acts as the information collection node of the Wearable Internet of Things (WIOT) framework in *EchoWear* [32]. Each participant wore an ASUS ZenWatch smartwatch paired to a corresponding Nexus 9 tablet. The data collection was initiated by the *EchoWear* application installed on an Android tablet. Once *EchoWear* was set in recording mode, the smartwatch collected the speech signal using the Android's Audio Application Program Interface (API.) The speech signal acquired was of high fidelity - 16-bit precision on sampling rate of 44.1 kHz. The smartwatch sends speech data to the tablet via Bluetooth 4.0. The tablet stores the speech data that can then be processed to remove anomalies. This processed speech signal was used to compute features for assessment of speech quality. Figure 1 shows the acoustic scenario used for capturing the mixed speech signals. It will be explained in later sections.

### B. Signal Processing Chain for Social Scenarios

The speech signal received from the tablet is analyzed by a series of signal processing methods to determine speech features. In addition to the analysis unit, it contains a display unit for displaying statistics of the corresponding speech data. The speech processing chain is shown in Figure 2. The mixed speech signal is the mixture of voices of several participants while doing vocal exercises or conversation in social scenarios such as group settings. The blind source separation (BSS) block takes the mixed speech signal and separates it into individual speech signals. The separated signals are fed into a noise reduction block where spectral subtraction is used for enhancing the noisy speech signal. This block outputs the enhanced speech signals that are used to compute speech features based on mathematical models (psychoacoustic models). The display unit visualizes the computed features that can be used by SLPs to monitor speech characteristics. Consider a scenario where three participants, using *EchoWear*, sit close to each other and simultaneously perform the vocal exercises. Each of the three smartwatches captures the mixed signals containing all three speech signals. Thus, the three mixed signals can be written as

$$x_1[n] = a_{11}s_1[n] + a_{12}s_2[n] + a_{13}s_3[n] \quad (1)$$
$$x_2[n] = a_{21}s_1[n] + a_{22}s_2[n] + a_{23}s_3[n] \quad (2)$$
$$x_3[n] = a_{31}s_1[n] + a_{32}s_2[n] + a_{33}s_3[n] \quad (3)$$

where $s_1[n]$, $s_2[n]$ and $s_3[n]$ are the speech signals from each of the participants, and $a_{ij}$'s for i, j = 1, 2, 3 are the mixing weights that depends on the relative distance between smartwatches and participants. We can only collect the mixed signals from such scenarios. Humans are keen at individualizing sound. However, for an algorithm it is difficult to get the source signals from the mixed ones. One such method is the independent component analysis (ICA). The separated speech signals $\hat{s}_1[n]$, $\hat{s}_2[n]$, and $\hat{s}_3[n]$ are obtained as linear transform of mixed signals. These estimates of source signals are not necessarily the true signals. ICA decomposes a set of mixed signals into statistically independent source signals. The source signals generating the mixture are assumed to be independent of each other.

ICA has been used in a variety of applications [19]-[23]. It can be used for feature extraction or blind signal separation (BSS). We have used ICA as signal separation algorithm for BSS in the speech processing sub-system. The estimation of individual signals $\hat{s}_1[n]$, $\hat{s}_2[n]$ and $\hat{s}_3[n]$ with the knowledge of mixed signals $x_1[n]$, $x_2[n]$ and $x_3[n]$ is known as the *cocktail-party problem* [24]. If we know the parameters $a_{11}, a_{12},…, a_{32}, a_{33}$ in Equations 1-3, then we can estimate the individual signals by solving the equations for $s_1[n]$, $s_2[n]$ and $s_3[n]$. However, we do not know the mixing weights in real scenarios, and due to this, the separation is challenging. There are two ambiguities in ICA, namely order ambiguity and variance ambiguity. Firstly, it is possible to estimate the independent component up to a scale i.e., the variance is not uniquely defined. We handle this issue by normalizing all separated signals to unit variance. This ensures speaker independent processing leading to accurate speech monitoring. Secondly, the order of components is not known exactly. This ambiguity is handled by subjective listening. We listen to the speech signals to identify the corresponding participant. Non-Gaussianity is the key quantity in the estimation of independent components. In accordance with central limit theorem, the sum of two independent random variables has a distribution that is closer to Gaussian than any of the two originals. For ICA estimation several quantitative measures of non-Gaussianity can be used e.g., kurtosis, neg-entropy. ICA can be estimated by maximization of non-

Gaussianity or minimization of mutual information. There are two key requirements for estimation of ICA model, finding an objective function and an algorithm for maximizing or minimizing it.

## III. METHOD

### A. Participants

The participants all had hypokinetic dysarthria secondary to Parkinson disease. They were recruited by the Department of Communicative Disorders at the University of Rhode Island. Recordings were obtained in a group setting with participants seated around rectangular table. There were 12 people in the group and three participants using *EchoWear* smartwatch system sat close to each other during a variety of vocal exercises and speech tasks. The acoustic scene for the group setting is depicted in Figure 1.

### B. Protocol and Data Collection

The data analyzed in this paper were collected in a group setting during a variety of vocal exercise and speech tasks. The participants were asked to perform four different tasks as shown in Table I. Tasks 1, 2, and 3 were performed by the participants simultaneously. Task 4 involved only one person speaking at a time. Thus, the smartwatch acquired a variety speech signals from individual speakers. The multi-smartwatch system proposed in this paper allows the participants to perform vocal exercises at home in small or large groups. The smartwatch can also be used to collect speech samples from everyday life in ecologically valid settings. There were two challenges for computation of speech features for each participant that were addressed in this study. Firstly, the smartwatch collects the mixture of speech and secondly, the background noise is present in acquired speech signals. It is important to develop a methodology for separation of mixed speech signal into individual signals and reduce the background noise

TABLE I. FOUR TASKS IN GROUP SETTINGS

| TASK 1 | The participants sustain the sound 'ahh' for as long as possible and repeat it three times. |
|---|---|
| TASK 2 | The participants start saying 'ahh' at their normal talking pitch and then go up in pitch. |
| TASK 3 | The participants start saying 'ahh' at normal pitch and then go down in pitch. |
| TASK 4 | The participants read sentences of 5-7 words in length. |

### C. Experimental Setup

The participants sit around a rectangular table along with the SLP leading the voice and speech exercises. These participants wore smartwatches that acquired the speech signal. This scenario is similar to a cocktail party problem that is a well-researched topic in speech processing literature [19]. In addition to mixing, each of the voices is contaminated with noise that can be approximated by additive white Gaussian noise. We used independent component analysis (ICA) for separating the mixture into component speech signals. The separated speech signals were processed with the spectral subtraction method for reducing the additive noise.

## IV. RESULTS AND DISCUSSIONS

In this section, we will discuss the results obtained using proposed method on clinical data. Task 4 did not need BSS for signal separation as only one person spoke at a time. In this paper, we will not discuss results for the task 4 as it involves a subset of processing needed for task 1, task 2 and task 3.

### A. Blind Separation of Speech Signals from Mixture

In this paper, we used Fast ICA for BSS that is a state-of-the-art ICA based BSS method [20]. Fast ICA requires the pre-processing of a mixed signal by centering (removing the mean) and whitening. It is based on a fixed-point iterative method for maximizing the non-Gaussianity in mixed signals. The Fast ICA algorithm estimates the mixing weights as well as the independent components. Kurtosis is a measure of non-Gaussianity and is defined by $kurt(Y) = E(Y^4) - 3$ for a random variable $Y$ such that $E(Y) = 0$ and $E(Y^2) = 1$. It is zero for zero-mean, unit variance Gaussian variables and is non-zero for most (but not all) non-Gaussian variables. Since kurtosis is sensitive to outliers in the measured data it is not suitable for use in Fast ICA algorithm. The entropy of a random variable $Y$ with probability density function $f_Y(y)$ is given by

$$H(Y) = -\int f_Y(y) \log(f_Y(y)) dy. \quad (4)$$

The Gaussian has the largest entropy among all random variables with same equal variance. Neg-entropy is an alternative measure of non-Gaussianity given by

$$J(Y) = H(Y_{gauss}) - H(Y), \quad (5)$$

where $Y_{gauss}$ is a multivariate Gaussian distribution with same covariance matrix as $Y$. Neg-entropy is always positive for non-Gaussian and zero for Gaussian distribution. It can be approximated as

$$J(Y) = k[E(G(Y)) - E(G(v))]^2 \quad (6),$$

where $v \geq 0$, is a Gaussian random variable with zero mean and unit variance. G in Equation 6 is a non-quadratic function used to avoid the problems due to outliers. Commonly used G are $G(u) = \tanh(u)$ and $G(u) = e^{-u^2/2}$. The observed mixed signal is pre-processed to remove the mean and normalize the variance such that $E(X) = 0$ and $E(XX^t) = I$. The pre-processed observed signal $X'$ is given by $X' = PD^{-\frac{1}{2}}P^t X$ where $E(XX^t) = PDP^t$ and $P^t P = I$ for a diagonal matrix $D$. The pre-processed mixed signals are further processed by fixed-point iteration of Fast ICA. Fast ICA finds successive vectors $w$ by solving following optimization problem

$$\max J(w^t x) \; s.t. \; var(w^t x) = 1, i.e., \quad \|w\| = 1. \quad (7)$$

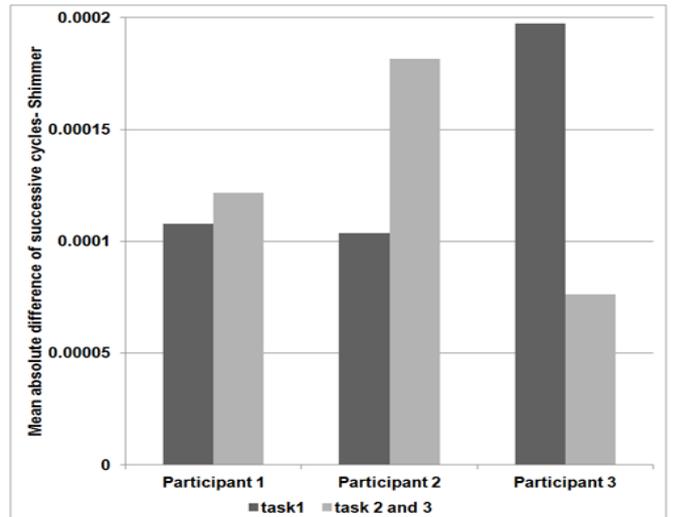

Fig. 3. Mean absolute difference of successive cycles in amplitude contour for various tasks.

The $w$ is orthogonal to all previously determined directions. In this way, we can get the required number of directions all at once. We decorrelated $w_1^t, w_2^t, \ldots, w_N^t$ after each iteration so that they converge to different maxima. It follows an approximate Newton iteration until a certain threshold is met. It achieves orthogonality using a simple Gram-Schmidt scheme in contrast to other ICA methods based on eigenvalue decomposition of a symmetric matrix.

It is possible to estimate the individual speech signals accurately using Fast ICA if and only if the original signals generating the mixture are statistically independent. This assumption is only partially valid for group settings. It leads to inherent inaccuracy in separated speech signals. However, these inaccuracies could not lead to perceptible changes in speech quality. This validates the applicability of Fast ICA for blind source separation of mixed speech signals obtained from social scenarios.

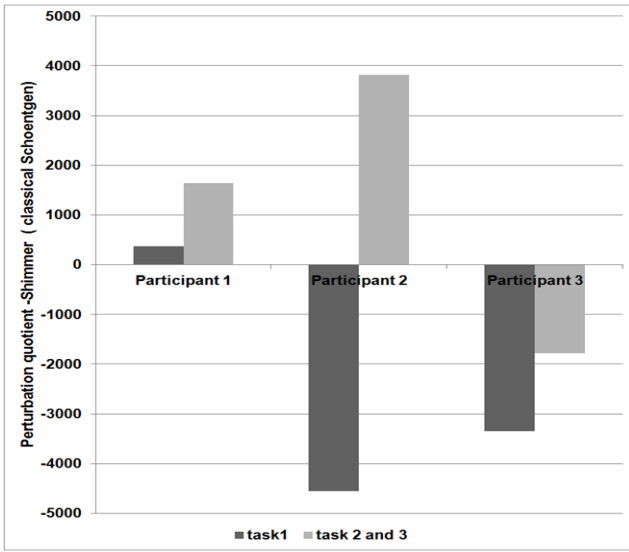

Fig. 4. Perturbation quotient (classical Schoentgen) of amplitude contour for various tasks.

### B. Noise Reduction by Spectral Subtraction

The separated speech signals were contaminated with noise. We used a computationally simple spectral subtraction technique for noise reduction [25]. The speech pauses are used to estimate the noise magnitude that was subtracted from the magnitude of noisy speech signal (separated signal). The phase of the noisy speech signal was used to reconstruct the enhanced speech signal. Inverse Fourier transform converts the enhanced speech spectrum into time-domain enhanced speech.

### C. Amplitude Variations: Shimmer

Shimmer is a measure of variation in amplitude of the speech signal. It is an important speech quality metric for people with speech disorders [26]-[27]. It aims to capture instabilities of the oscillating pattern of the vocal folds that influences the cycle-to-cycle changes in amplitude. The speech disorders are due to imprecise movement of vocal folds. The method presented in [28] is used to compute shimmer for enhanced speech signal. In total there are 22 features to characterize shimmer. We computed all 22 features but showing only two features in Figure 3 and 4 that are mean absolute difference of successive cycles, and perturbation quotient (classical Schoentgen) respectively for amplitude contour. It is clear from these figures that shimmer features can be used for quantification of perceptual speech quality.

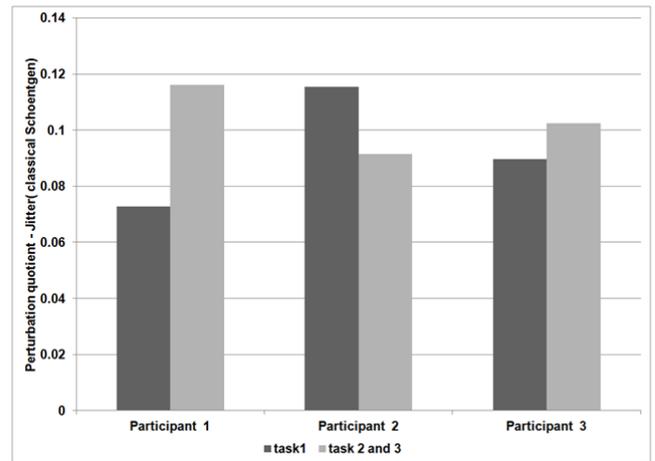

Fig. 6. Perturbation quotient (classical Schoentgen) of pitch contour for various tasks.

### D. Pitch Variations: Jitter

Speech disorders lead to monotonous pitch with imprecise articulation. The fundamental frequency ($F_0$) in Hz was estimated by Praat's time-domain approach based on autocorrelation [29]. Praat was used to divide the short-time stationary speech signal into stationary segments. Speech was divided into overlapping frames of size 40-80 millisecond (ms) so that each frame is almost stationary. The mean value is subtracted from each sample in that speech segment followed by time domain multiplication with the Hanning window. This gives the windowed speech segments in time-domain. The time-period was calculated from each of the windowed segments by taking the mid-sample within each window and finding the most correlated time-instant. The normalized autocorrelation function is obtained by dividing the signal autocorrelation by autocorrelation of the Hanning window. The number of samples between peaks in the normalized autocorrelation function of the windowed speech signal was calculated. The fundamental frequency was given by sampling frequency divided by the number of samples between autocorrelation peaks. Jitter measures the time variability of the pitch. The fundamental frequency contour was used to obtain the jitter features derived in [28]. Similar to shimmer there are total of 22 features to characterize jitter. We computed all 22 features but showing only two features in Figure 5 and 6 that are mean absolute difference of successive cycles, and perturbation quotient (classical Schoentgen) respectively for pitch contour. The shimmer and jitter features are statistical quantities capturing cycle-to-cycle variability in speech amplitude and pitch respectively. The pitch in semitone is given by [30]

$$p = 69 + 12 \log_2 \left(\frac{F_0}{440}\right) \qquad (8)$$

where $F_0$ is the fundamental frequency (in Hz). The semitone standard deviation, that is standard deviation of $p$ contour, is used as a measure of pitch variability [31]. The semitone standard deviation is shown in Figure 7 for all separated speech signals. It can be seen that task 2 and task 3 were combined into a single audio files for the purposes of blind source separation followed by computation of speech quality analytics. Combining task 2 (HIGHS) and task 3 (LOWS) gives an audio files that records both task one after the other. The resulting audio files were longer in duration and more stable for computing a sufficient number of independent components that were equal to the number of participants (with smartwatches). Figure 7 clearly shows that semitone standard deviation captured the variability in pitch of the speech signal. High semitone standard deviation reflects higher pitch variability that means better speech quality. For example, participant 2 had better speech quality than participant 1 with respect to task 1.

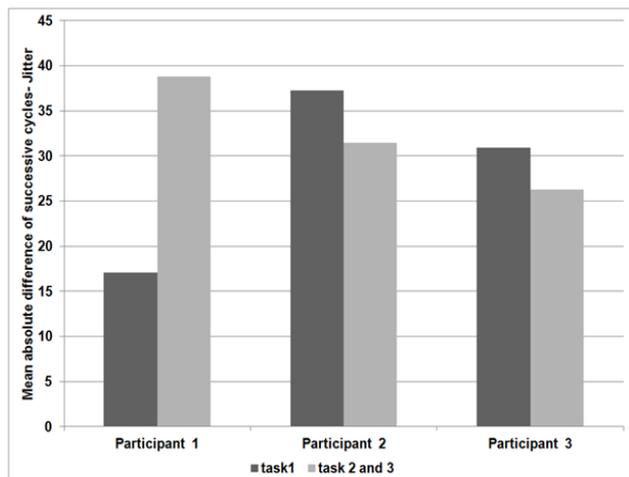

Fig. 5. Mean absolute difference of successive cycles in pitch contour for various tasks.

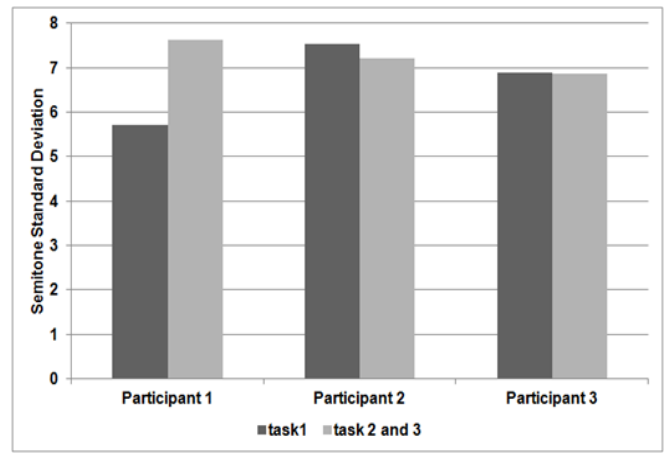

Fig. 7. Semitone standard deviation for various tasks.

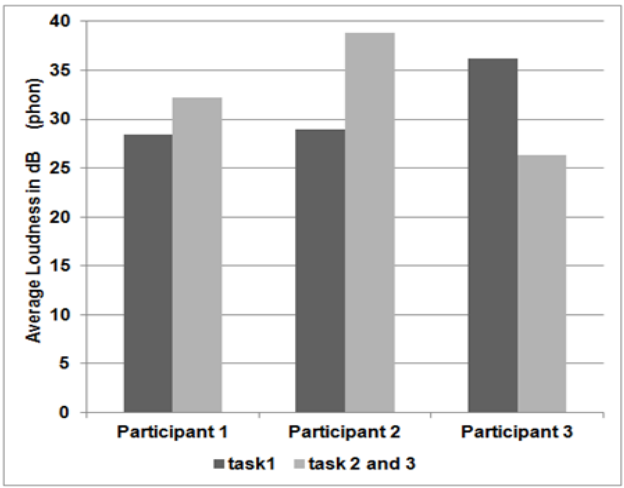

Fig. 8. Average loudness in dB (Phon) unit for three participants on various tasks.

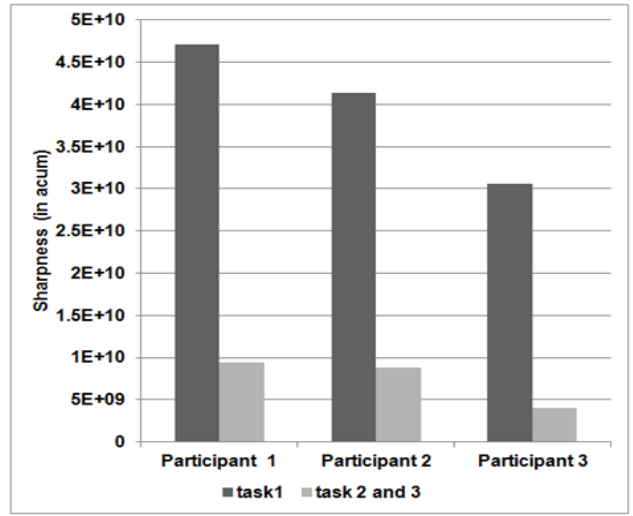

Fig. 9. Sharpness (in acum) for all participants with respect to various tasks.

*E. Loudness*

We previously computed the loudness of speech signals using Zwicker's method for time varying signals [18]. Using the same method on enhanced speech signals we get the instantaneous loudness contour. Figure 8 shows the mean loudness in dB (Phon) that was obtained by averaging the loudness contour for each participant. It can be clearly seen that the loudness for speaker participant 3 was higher than both participant 1 and participant 2 for task 1. Loudness is responsible for intensity sensation in human auditory system and depends on amplitude, time duration and frequency content of speech signal [33].

*F. Sharpness*

Sharpness is related to sensory pleasantness of a voice. High sharpness implies low sensory pleasantness. Sharpness depends on other sensations such as loudness and tonalness. The sensation of sharpness is influenced by the spectral envelope of speech signal. The spectrally fine structure is not important for sharpness. Sharpness depends on the level and the bandwidth of sound. The Latin expression 'acum' is used as unit for sharpness [33]. Figure 9 shows the sharpness (in acum) for each enhanced speech signal. It is clear that the participant 1 has highest sharpness i.e., lowest sensory pleasantness for task 1. However, participant 3 has higher sensory pleasantness for task 1 as compared to participant 1 and participant 2.

V. CONCLUSION

This paper extends the reliability and validity of the *EchoWear* system to measure characteristics of speech sampled in a sound booth to a multi-smartwatch system in which speech was sampled in a group setting – providing a more representative ecological valid environment for collecting speech data. It brings the application of blind source separation to the group settings where participants simultaneously performed vocal exercises and spoke individually in a group setting. The data support the validity of the system and methods described to analyze voice and speech characteristics of an individual when multiple speakers are present. Due to practical limitations small deviations are inevitable. However such deviations were insignificant for the purpose of speech monitoring. The proposed method appears valid for speech monitoring during home exercises and in ecologically valid social scenarios, such as in the context of a meal with family. Future research studies will examine new analytics or metrics to allow the SLP to access speech and voice data while the patient uses the device outside of the clinical setting for assessment and monitoring speech production outside the clinical setting. Deploying remote trails under varied acoustic scenarios, e.g., adverse situations such as high background noise, reverberation etc. is another future direction for assessing the validity and reliability of the *EchoWear* system. This technology has the potential to be useful for SLPs as well as patients with dysarthria by providing a method for speech monitoring during individual exercise and communication in a group setting. Therefore, it has the potential to be an important adjunct to treatment for carryover and generalization of speech goals to social, ecologically valid scenarios.

ACKNOWLEDGMENT

This research is supported by Rhode Island Foundation (grant no. 20144261).